\documentclass[10pt]{article}
\usepackage{amsmath}
\usepackage[OE]{express}
\usepackage[framed,numbered,autolinebreaks,useliterate]{mcode}

\begin{document}

\title{Numerical realization of the variational method for generating
self-trapped beams}
\author{Erick I. Duque,\authormark{1} Servando Lopez-Aguayo,%
\authormark{1,*} and Boris A. Malomed\authormark{2,3}}

\address{\authormark{1} Tecnologico de Monterrey, Escuela de Ingenier\'ia
y Ciencias, Ave. Eugenio Garza Sada 2501, Monterrey, N.L., M\'exico, 64849.}

\address{\authormark{2} Department of Physical Electronics, School of
Electrical Engineering, Faculty of Engineering and Tel Aviv University
Center for Light-Matter Interaction, Tel Aviv 69978, Israel}

\address{\authormark{3} ITMO University, St. Petersburg 197101, Russia}

\email{\authormark{*}servando@itesm.mx}

\begin{abstract}
We introduce a numerical variational method based on the Rayleigh-Ritz
optimization principle for predicting two-dimensional self-trapped beams in
nonlinear media. This technique overcomes the limitation of the traditional
variational approximation in performing analytical Lagrangian integration
and differentiation. Approximate soliton solutions of a generalized
nonlinear Schr\"{o}dinger equation are obtained, demonstrating robustness of
the beams of various types (fundamental, vortices, multipoles, azimuthons)
in the course of their propagation. The algorithm offers possibilities to
produce more sophisticated soliton profiles in general nonlinear models.
\end{abstract}

% REPLACE WITH CORRECT OCIS CODES FOR YOUR ARTICLE, MINIMUM OF TWO; Avoid using the OCIS codes for “General” or “General science” whenever possible.
%For a complete list of OCIS codes, visit: https://www.osapublishing.org/oe/submit/ocis/

%%%%%%%%%%%%%%%%%%%%%%% References %%%%%%%%%%%%%%%%%%%%%%%%%
% \begin{thebibliography}{99}
% \bibitem{Anderson} D. Anderson, M. Lisak, \& A. Berntson, ``A variational approach to nonlinear evolution equations in optics,'' J. Phys., Vol. {\bfseries 57}(5 \& 6), 917--936 (2001).
% \bibitem{Mihalache} D. Mihalache, D. Mazilu, et. al. ``Stable spatiotemporal solitons in bessel optical lattices,'' Physical Review Letters, vol. {\bfseries 95}(2), 10703--10710 (2005).
% \bibitem{Nonlocal_variational_1D} Dai, Z., Yang, Z., Ling, X., Zhang, S., Pang, Z., \& Li, X. (2017). Tripole-mode and quadrupole-mode solitons in media with a spatial exponential- decay nonlocality. Scientific Reports, (February), 1–13.
% \bibitem{Perturbed_Variational} Sahoo, A., Roy, S., \& Agrawal, G. P. (2017). Perturbed dissipative solitons: A variational approach. Physical Review A, 96(1), 1–8.

% \bibitem{MATLAB_code} Algorithm implemented in MATLAB.

% \end{thebibliography}

\bibliographystyle{osajnl}
\bibliography{references.bib}

%%%%%%%%%%%%%%%%%%%%%%%%%%  body  %%%%%%%%%%%%%%%%%%%%%%%%%%

\section{Introduction}

% The Rayleigh-Ritz Optimization is considered an analytical approximation to the solution of the nonlinear Schr\"{o}dinger equation (NLSE) as we know exactly the shape of $\psi$. Nevertheless, the big disadvantage of this method is that the integrations to get the effective Lagrangian are often difficult for more complicated trial functions and richer nonlinear potentials. Furthermore, even if the integration is analytically possible, the resulting equations of the second variation may need to be solved numerically anyway.\\
Self-trapped beams in nonlinear media, alias optical spatial solitons, have
long been the subject of intensive studies, promising new physical effects
and potential application to optic-communication systems and photonic
technologies \cite{yuribook,chen}. Common theoretical models for producing
optical solitons rely on the use of generalized nonlinear Schr\"{o}dinger
equations (GNLSEs), which usually do not admit exact solutions. Analytical
techniques, such as the inverse scattering method \cite{inverse} and B\"{a}%
cklund transform \cite{back}, apply only to exceptional integrable systems.
Numerical methods for generating soliton solutions, including the standard
relaxation algorithm and conjugate gradient method \cite{yang} are very
useful, but they do not always provide sufficient insight.
In particular, an iterative method was developed for finding numerical eigenfunctions and eigenvalues
corresponding to soliton solutions of the nonlinear Schr\"{o}dinger equation,
and applied it in a variety of cases (see \cite{Matusevich2008, Matusevich2009}). Similar to the situation with the standard relaxation algorithm, the success of the iterative method depends on the appropriate choice of the initial guess \cite{Trofimov2010}.
To provide deeper understanding of nonlinear systems, approximate semi-analytical methods have
been developed, the most useful one being, arguably, the variational
approximation (VA) based on the Rayleigh-Ritz optimization of a trial
function, alias \textit{ansatz}, that was introduced in the context of pulse
propagation in nonlinear fiber optics in the seminal paper by Anderson \cite%
{anderson} (see also works \cite{bondeson} and \cite{reichel}). The VA makes
it possible to predict nonlinear modes with different degrees of accuracy,
depending on the choice of the underlying \textit{ansatz} \cite{boris}, in
diverse nonlinear systems. These include, in particular, photonic lattices
in nonlinear media, induced by non-diffracting beams \cite{kartashov},
dissipative media \cite{skarka,Sahoo2017}, Bose-Einstein condensates \cite%
{bec,Mihalache2005}, parity-time-symmetric lattices \cite{Hu2014}, dynamics of spatio-temporal solitons in a periodic medium \cite{Aceves1993}, and even
the prediction of complex \textit{hopfion} modes with two topological
charges (twisted vortex ring) \cite{hopfion}. The VA was also applied to
nonlocal GNLSEs \cite{konotop,Dai2017}. Furthermore, predicted the approximate solutions predicted by the VA can be used in the context of the above-mentioned numerical methods, as an appropriate initial distribution which accelerates the convergence to numerically exact solutions.

Coming along with assets of the VA are its technical limitations, as in many
cases it may be problematic to perform analytical integration of the
underlying Lagrangian with the substitution of the chosen ansatz, and
subsequent differentiation with respect to free parameters, aiming to derive
the Euler-Lagrange (EL)\ equations. In this work, we present an algorithm
that permits a full numerical treatment of the VA based on the Rayleigh-Ritz
optimization, overcoming its intrinsic limitations. In fact, the development
of such fully numerical approach is natural, as, in most cases, the EL
equations are solved numerically, anyway. In particular, we use the
numerically implemented VA to obtain approximate solutions for rotary multipole modes, as
well as vorticity-carrying ring-shaped solitons, in the context of the GNLSEs with various nonlinear terms, cf. Ref. \cite{hopfion}.

\section{Theoretical framework}

\subsection{Lagrangian}

Evolution equations for complex amplitude $\Psi $ of the optical field are
derived from action
\begin{equation}
S=\int \int \mathcal{L}{d}t{d}\mathbf{r}=\int L{d}t,
\end{equation}%
where $t$ is the evolution variable, $\mathbf{r}$ a set of transverse
coordinates, $\mathcal{L}$ the Lagrangian density and $L=\int \mathcal{L}d%
\mathbf{r}$ the full Lagrangian. As we search for soliton solutions, the field and its derivatives must vanish at boundaries of the integration domain, which emulate the spatial infinity in terms of the numerical solution. The respective EL equation follows from the
action-minimum principle, $\delta S/\delta \Psi ^{\ast }=0$, i.e.,%
\begin{equation}
\frac{d}{dt}\left( \frac{\partial \mathcal{L}}{\partial \Psi _{t}^{\ast }}%
\right) +\sum_{r_{i}=x,y,z}\frac{d}{dr_{i}}\left( \frac{\partial \mathcal{L}%
}{\partial \Psi _{r_{i}}^{\ast }}\right) -\frac{\partial \mathcal{L}}{%
\partial \Psi ^{\ast }}=0,  \label{eq:func_deriv}
\end{equation}%
where $\Psi ^{\ast }$ is the complex conjugate $\Psi $.
In particular, the GNLSE, in the form of
\begin{equation}
i\partial _{t}\Psi +\nabla ^{2}\Psi +N(|\Psi |^{2},\mathbf{r})\Psi =0,
\label{eq:GNLSE}
\end{equation}%
with the nonlinear part $N(|\Psi |^{2},\mathbf{r})\Psi $ \cite{Anderson2001}%
, is generated by
\begin{equation}
\mathcal{L}=\frac{i}{2}(\Psi \partial _{t}\Psi ^{\ast }-\Psi ^{\ast
}\partial _{t}\Psi )+|\nabla \Psi |^{2}+\mathcal{NL}(|\Psi |^{2},\mathbf{r}),
\label{eq:L_GNLSE}
\end{equation}%
where $\mathcal{NL}(|\Psi |^{2},\mathbf{r})$ is the \textit{anharmonic} term
in the Lagrangian density which gives rise to $N(|\Psi |^{2},\mathbf{r})\Psi
$ in the GNLSE.% Noting that

\subsection{Numerical variational method}

% The integral to get $L_{\textbf{(A)}}$ is often difficult to perform analytically for complicated Ansatz and nonlinear potentials. The following algorithm calculates the optimal coefficients of an Ansatz to satisfy the equation with no need to solve $L_{\textbf{(A)}}$ analytically. Only the expression of the Ansatz $\psi$ in a standard Lagrangian form must be provided to the algorithm.\\
% The optimization of $\textbf{A}$ occurs when $\nabla_\textbf{A} L_{(\textbf{A})}=0$ is sattisfied.
%If $L_{\textbf{(A)}}$ is known, everything left to do is to satisfy the optimization $\nabla_\textbf{A} L_{(\textbf{A})}=0$.
We start by defining vectors $\mathbf{A}=(A_{1},A_{2},...,A_{n})$ and $%
\nabla _{\mathbf{A}}L_{\mathbf{(A)}}=(\partial _{A_{1}}L_{\mathbf{(A)}%
},\partial _{A_{2}}L_{\mathbf{(A)}},...,\partial _{A_{n}}L_{\mathbf{(A)}})$,
where $A_{1}$, $A_{2}$,..., $A_{n}$ are variational parameters, such as an
amplitude, beam's width, frequency chirp, etc. We adopt ansatz $\Psi _{(%
\mathbf{r},\mathbf{A})}$, followed by the integration of $\mathcal{L}$ to
produce the respective effective Lagrangian, $L_{\mathbf{(A)}}$, as a
function of variables $A_{n}$. As we aim to find solutions for self-trapped beams, we adopt zero boundary conditions at boundaries of the numerical-integration domain.
The variation of $L_{\mathbf{(A)}}$ gives
rise to the system of the EL equations:
\begin{equation}
\frac{d}{dt}\left( \frac{\partial L_{\mathbf{(A)}}}{\partial \left(
dA_{n}/dt\right) }\right) -\frac{\partial L_{\mathbf{(A)}}}{\partial A_{n}}%
=0\ ,
\label{eq:EL_equation}
\end{equation}%
whose stationary version can be written as $\nabla _{\mathbf{A}}L_{(\mathbf{A%
})}=0$, with $\nabla _{\mathbf{A}}$ standing for the set of the derivatives
with respect to $A_{n}$. Generally, the stationary EL\ equations cannot be
solved analytically, therefore we apply the standard Newton-Raphson
multi-variable method, with the Jacobian replaced by the Hessian matrix,
% \begin{equation}
% H_{L_{(\textbf{A})}} =
% \begin{bmatrix}
% \frac{\partial^2 L_{\textbf{(A)}}}{{\partial A_1}^2} & \frac{\partial^2 L_{\textbf{(A)}}}{\partial A_2 \partial A_1} & \dots & \frac{\partial^2 L_{\textbf{(A)}}}{\partial A_n \partial A_1} \\
% \frac{\partial^2 L_{\textbf{(A)}}}{\partial A_1 \partial A_2} & \frac{\partial^2 L_{\textbf{(A)}}}{{\partial A_2}^2} & \dots & \frac{\partial^2 L_{\textbf{(A)}}}{\partial A_n \partial A_2}\\
% \dots & \dots & \dots & \dots \\
% \frac{\partial^2 L_{\textbf{(A)}}}{\partial A_1 \partial A_n} & \frac{\partial^2 L_{\textbf{(A)}}}{\partial A_2 \partial A_n} & \dots & \frac{\partial^2 L_{\textbf{(A)}}}{{\partial A_n}^2}
% \end{bmatrix}.
% \end{equation}
\begin{equation}
HL_{(\mathbf{A})}=%
\begin{bmatrix}
\partial^2_{A_1} L_{\mathbf{(A)}} & \partial_{A_2} \partial_{A_1} L_{\mathbf{%
(A)}} & \dots & \partial_{A_n} \partial_{A_1} L_{\mathbf{(A)}} \\
\partial_{A_1} \partial_{A_2} L_{\mathbf{(A)}} & \partial^2_{A_2} L_{\mathbf{%
(A)}} & \dots & \partial_{A_n} \partial_{A_2} L_{\mathbf{(A)}} \\
\dots & \dots & \dots & \dots \\
\partial_{A_1} \partial_{A_n} L_{\mathbf{(A)}} & \partial_{A_2}
\partial_{A_n} L_{\mathbf{(A)}} & \dots & \partial^2_{A_n} L_{\mathbf{(A)}}%
\end{bmatrix}%
.
\end{equation}%
In the framework of this method, an the iterative search for the solution is
carried out as
\begin{equation}
\mathbf{A}^{(i+1)}=\mathbf{A}^{(i)}-HL_{\mathbf{(A)}}^{-1}\cdot \nabla _{%
\mathbf{A}}L_{(\mathbf{A})},  \label{eq:iterate}
\end{equation}%
starting with an initial guess.

However, the substitution of necessarily complex \textit{ans\"{a}tze }in
Lagrangians of many nonlinear models leads to analytically intractable
integrals. Thus, neither $L_{\mathbf{(A)}}$ nor $\nabla _{\mathbf{A}}L_{%
\mathbf{(A)}}$ or $HL_{\mathbf{(A)}}$ may be known in an explicit analytical
form. This difficulty is exacerbated by working with multidimensional
settings and using, if necessary, involved coordinate systems. In this work,
we develop a way to overcome this limitation, noting that, to produce $%
\nabla _{\mathbf{A}}L_{\mathbf{(A)}}$ and $HL_{\mathbf{(A)}}$, which are
needed to apply the Newton-Raphson method, one can numerically calculate $%
L_{(\mathbf{A})}$ at multiple points in the space of variables $A_{n}$,
separated by small finite values $\Delta A_{n}$. In particular, the
derivatives can be computed as
\begin{equation}
\frac{\partial L_{\mathbf{(A)}}}{\partial A_{n}}=\int \frac{\partial
\mathcal{L_{(\mathbf{A})}}}{\partial A_{n}}d\mathbf{r}\approx \int \frac{%
\mathcal{L}_{(\mathbf{A}+\Delta A_{n})}-\mathcal{L}_{(\mathbf{A}-\Delta
A_{n})}}{2\Delta A_{n}}d\mathbf{r}=\frac{L_{(\mathbf{A}+\Delta A_{n})}-L_{(%
\mathbf{A}-\Delta A_{n})}}{2\Delta A_{n}},  \label{First_derivative}
\end{equation}%
%
%
%
% \begin{equation}
% \partial L_{\textbf{(A)}}/\partial A_n \approx (L_{(\textbf{A} + \Delta A_n)} - L_{(\textbf{A} - \Delta A_n)})/2 \Delta A_n,
% \label{First_derivative}
% \end{equation}
%this is the simplest way of calculating such derivative numerically. We calculate likewise,
and, similarly,%
\begin{equation}
\frac{\partial ^{2}L_{\mathbf{(A)}}}{\partial A_{n}\partial A_{m}}\approx
\frac{L_{(\mathbf{A}+\Delta A_{n}/2+\Delta A_{m}/2)}-L_{(\mathbf{A}-\Delta
A_{n}/2+\Delta A_{m}/2)}-L_{(\mathbf{A}+\Delta A_{n}/2-\Delta A_{m}/2)}+L_{(%
\mathbf{A}-\Delta A_{n}/2-\Delta A_{m}/2)}}{\Delta A_{n}\Delta A_{m}}.
\label{Double_derivative}
\end{equation}%
%
%
%
% \begin{equation}
% \frac{\partial^2 L_{\textbf{(A)}}}{\partial A_n \partial A_m} \approx \frac{L_{(\textbf{A} + \frac{\Delta A_n}{2} + \frac{\Delta A_m}{2})} - L_{(\textbf{A} - \frac{\Delta A_n}{2} + \frac{\Delta A_m}{2})} - L_{(\textbf{A}+ \frac{\Delta A_n}{2} - \frac{\Delta A_m}{2})} + L_{(\textbf{A} - \frac{\Delta A_n}{2} - \frac{\Delta A_m}{2})}}{\Delta A_n \Delta A_m}.
% \label{Double_derivative}
% \end{equation}
% which for $n=m$ reduces to the standard one variable double finite difference derivative
% \begin{equation}
% \frac{\partial^2 L_{\textbf{(A)}}}{{\partial A_n}^2} \approx \frac{L_{(\textbf{A} + \Delta A_n)} - 2L_{\textbf{(A)}} + L_{(\textbf{A} - \Delta A_n)}}{{\Delta A_n}^2}.
% \end{equation}
Note that each step in this procedure can be done numerically. While there
is freedom in the choice of $\Delta A_{n}$, it is reasonable to select their
size smaller than an estimated solution for $A_{n}$ by at least three orders
of magnitude. Thus, the algorithm can be summarized as follows:

\begin{itemize}
%\item
%\item Select any initial combination $\textbf{A}$ of the coefficients.

\item Calculate $\nabla _{\mathbf{A}}L_{\mathbf{(A)}}$ and $HL_{\mathbf{(A)}%
} $ numerically with the help of Eqs. (\ref{First_derivative}) and (\ref%
{Double_derivative}), respectively. Because $HL_{\mathbf{(A)}}$ is a $%
n\times n$ symmetric matrix, only $n(n+1)/2$ different elements need to be actually
calculated.

\item Compute new $\mathbf{A}$ according to Eq. (\ref{eq:iterate}).

\item Iterate the two previous steps until achieving specified tolerance for
the convergence of $\mathbf{A}$.
\end{itemize}

A disadvantage of this algorithm is that the trivial zero solution also
satisfies the optimization procedure, hence the iterations may converge to
zero in some cases. A simple but effective way to avoid this is to select a
new starting point with larger values of $A_{n}$.
It is also worthy to note that, in the case of multistability, the algorithm can find different coexisting solutions, the convergence to a specific one depending on the choice of the initial guess.

\section{Numerical results for self-trapped modes}

Here we report results obtained for the GNLSEs generated by the Lagrangian
with anharmonic terms displayed in Table \ref{tab:Nonlinear_potentials}. We
start by looking for solutions for spatial solitons, in the form of
\begin{equation}
\Psi (x,y,z)=\psi (\mathbf{r})e^{i\lambda z},  \label{eq:lambda}
\end{equation}%
where $\mathbf{r}=\left\{ x,y\right\} $ are the transverse spatial
coordinates, longitudinal coordinate $z$ is the evolution variable, which
replaces $t$ in the outline of the variational method presented above, and $%
\lambda $ is a real propagation constant. The substitution of this waveform
simplifies GNLSE (\ref{eq:GNLSE}) and Lagrangian (\ref{eq:L_GNLSE}),
\begin{equation}
-\lambda \psi (\mathbf{r})+\nabla _{\perp }^{2}\psi (\mathbf{r})+N(|\psi
|^{2},\mathbf{r}) \psi (\textbf{r}) =0,  \label{eq:red_GNLSE}
\end{equation}%
\begin{equation}
\mathcal{L}=\lambda |\psi |^{2}+|\nabla _{\perp }\psi |^{2}+\mathcal{NL}%
(|\psi |^{2},\mathbf{r}).  \label{eq:Lagrangian_density}
\end{equation}

The iterative procedure for the solution of the optimization problem does not provide the
conservation of the dynamical invariants, \textit{viz}., the norm and Hamiltonian. However, this fact does
not invalidate the applicability of the procedure, similarly to the well-known imaginary-time
integration method, which provides correct stationary solutions, although the Hamiltonian is not
conserved in the course of the solution \cite{imaginary1,imaginary2} .

\begin{figure}[!htb]
\centering
\includegraphics[trim={0cm 5cm 0cm
5cm},clip,width=1\textwidth]{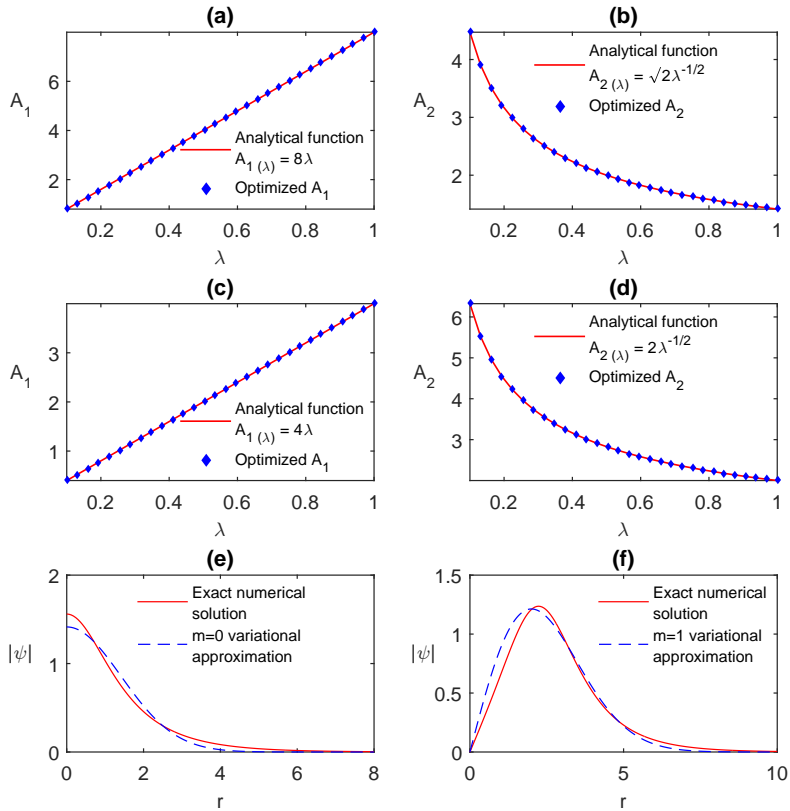}
\caption{The amplitude (a) and beam's width (b), predicted by the numerical
variational procedure for the fundamental soliton ($m=0$), compared to the
results of the analytical variational approximation (see the text). The
maximum relative differences for parameters $A_{1}$ and $A_{2}$ are $%
e_{A_{1}}=1.72\times 10^{-8}\ \%$ and $e_{A_{2}}=2.40\times 10^{-8}\ \%$ .
The same comparison for the amplitude (c) and width (d) of the vortex beam
with $m=1$ in the Kerr medium, with maximum relative differences $%
e_{A_{1}}=2.17\times 10^{-7}\ \%$ and $e_{A_{2}}=1.65\times 10^{-7}\ \%$. Comparison between the exact soliton shape and the variationally obtained using Eq. (\ref{eq:iterate}) for fundamental (e) and single vortex soliton (f), in both cases $\lambda=0.5$.}%
\label{fig:Kerr_Vortex_m0_m1_fit}
\end{figure}

In this work, we focus on the optimization of the following generalized
vortex ansatz with integer topological charge $m$ in two-dimensional GNLSEs
including anharmonic terms listed in Table \ref{tab:Nonlinear_potentials}:
\begin{gather}
\psi (\mathbf{r})=A_{1}^{(m+1)/2}\exp \left[ -\left( x/A_{2}\right)
^{2}-\left( y/A_{3}\right) ^{2}\right]  \notag \\
\times\left[ (\cos ^{2}\epsilon )~x^{2}+(\sin ^{2}\epsilon )~y^{2}\right]%
^{m/2}\left[(\cos \delta )\cos (m\theta )+i~\left( \sin \delta \right) \sin
(m\theta )\right] ,  \label{Gen_Trial_function}
\end{gather}%
$\theta \equiv \arctan (y/x)$ being the angular coordinate. The ansatz may
be construed as either an asymmetric \textit{azimuthon} \cite{azimut} or
\textit{ellipticon} \cite{servando}. Here, $A_{1}$ represents the amplitude
of the anisotropic beam, while $A_{2}$ and $A_{3}$ determine its width, $%
A_{2}=A_{3}$ corresponding to the isotropic one. Parameters $\epsilon $ and $%
\delta $ additionally control the beam's asymmetry and its phase structure.
In particular, for $\epsilon =\delta =\pi /4$ the ansatz reduces to a
standard vortex beam, while at $\delta =0$ it is reduced to a multi-pole
beam.

Generic results can be adequately presented for propagation constant $%
\lambda =1/2$ in Eq. (\ref{eq:lambda}) and values of parameters $s=0.05$ and $%
k=5$, $p=30$ in Lagrangian terms (\ref{eq:Saturation}) and (\ref{eq:Bessel}%
), respectively, unless specified otherwise.

First, we reproduce known results produced by the VA for the fundamental and
first-order vortex solitons in the Kerr medium, corresponding to Eq. (\ref%
{eq:Kerr}), based on normalized ansatz $\psi
=2^{-(m+1)/2}A_{1}^{(m+1)/2}\exp \left[ -(r/A_{2})^{2}\right] r^{m}\exp
(im\theta )$ \cite{Dimitrevski}. With this ansatz, the analytical VA yields $%
A_{1}=8\lambda $ and $A_{2}=\sqrt{2/\lambda }$ for $m=0$, and $%
A_{1}=4\lambda $ and $A_{2}=2/\sqrt{\lambda }$ for $m=1$. Using the present
algorithm leads to complete agreement with these results, as shown in Fig. %
\ref{fig:Kerr_Vortex_m0_m1_fit}, with relative errors $<3\times 10^{-7}\%$.
Each of these sets of the results can be generated by a standard computer in
less than a minute.
The general agreement between the exact soliton shape and the solution obtained by using the numerical variational approach is good, as is evident from Fig. \ref{fig:Kerr_Vortex_m0_m1_fit}(e) and Fig. \ref{fig:Kerr_Vortex_m0_m1_fit}(f) for the fundamental and single vortex soliton, respectively. In principle, more sophisticated ans\"{a}tze may provide closer agreement with numerically exact results, but in this work we focus on the basic results produced by the relatively simple ansatz.

Next, we proceed to demonstrate the utility of the algorithm for more
complex nonlinear media. We use the combination of the Kerr and first-order
Bessel-lattice terms, that is, the sum of terms given \ by Eqs. (\ref%
{eq:Kerr}) and (\ref{eq:Bessel}) . We test the utility of the algorithm by
comparing its predictions with simulations, using the standard split-step
Fourier-transform algorithm and presenting the evolution of the intensity
profile, $|\psi |^{2}$, for a certain size of the transverse display area
(TDA).

\begin{figure}[!htb]
\centering
\includegraphics[trim={5.5cm 12.5cm 4.25cm
9cm},clip,width=0.8\textwidth]{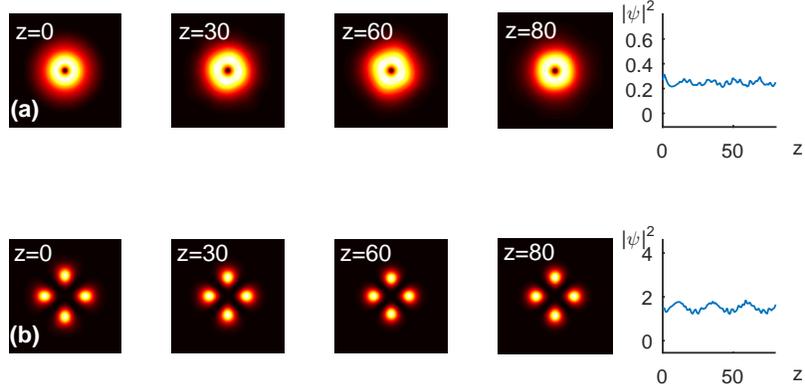}
\caption{Simulated propagation in the medium with the combined nonlinearity
corresponding to Eq. (\protect\ref{eq:Kerr}) + Eq. (\protect\ref{eq:Bessel}%
), initiated by inputs produced by the numerically implemented variational
approximation. (a) The vortex soliton with topological charge $m=1$ with $%
\mathbf{A_{1,2,3}}=(0.7830,2.7903,2.7903)$ and TDA $15\times 15$. (b) The
propagation of the quadrupole soliton in the medium with the combined
nonlinearity corresponding to Eq. (\protect\ref{eq:Kerr}) + Eq. (\protect\ref%
{eq:Bessel}), with $\mathbf{A_{1,2,3}}=(0.5767,3.4560,3.4560)$ and TDA $20\times 20$%
. The peak-intensity evolution is shown in the right-hand column.}
\label{fig:Saturation_KB1}
\end{figure}

Figure \ref{fig:Saturation_KB1} (a) displays the numerically simulated
propagation of the isotropic vortex soliton with $m=1$, starting from the
input predicted by the numerically implemented VA. Note that, while the
direct simulation does not preserve a completely invariant shape of the
vortex beam, the VA provides a reasonable prediction. It is relevant to
mention that the simulations do not include initially imposed azimuthal
perturbations, which may lead to breakup of the vortex ring due the
azimuthal instability \cite{review}.

\begin{figure}[!htb]
\centering
\includegraphics[trim={5.5cm 9.5cm 4.25cm
9cm},clip,width=0.8\textwidth]{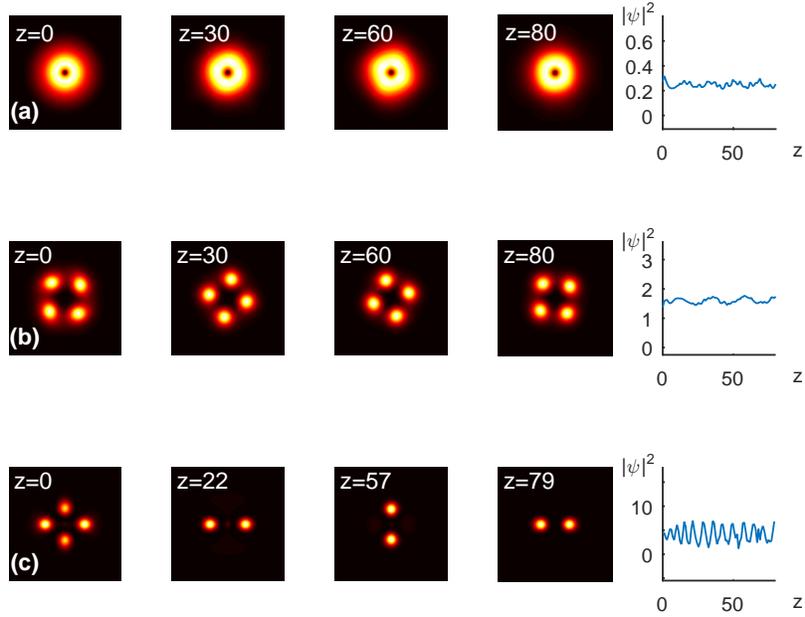}
\caption{The same as in Fig. \protect\ref{fig:Saturation_KB1}, but under the
action of the combined nonlinearity given by Eq. (\protect\ref{eq:Saturation}%
) + Eq. (\protect\ref{eq:Bessel}). (a) The vortex with $m=1$, $%
\mathbf{A_{1,2,3}}=(0.7860,2.7915,2.7915)$, and TDA $15\times 15$. (b) The azimuthon
with $m=2$, $\protect\delta =\protect\pi /3$, $%
\mathbf{A_{1,2,3}}=(0.63574,3.4760,3.4760)$, and TDA $20\times 20$. (c) The
asymmetric quadrupole with $s=0.2$, $\protect\lambda =0.8$, $\protect%
\epsilon =5\protect\pi /16$, $\mathbf{A_{1,2,3}}=(1.0334,3.4235,2.4701)$, and TDA $%
20\times 20$.}
\label{fig:Saturation_Bessel1}
\end{figure}

Then, we optimize a quadrupole soliton beam. The simulated propagation is
displayed in Fig. \ref{fig:Saturation_KB1}(b), where the persistence of the
VA-predicted shape is obvious. Note that the usual form of the VA cannot be
applied in this case, because of its complexity. Further, we proceed to
self-trapped beams supported by a combination of the saturable nonlinearity
and the first-order Bessel-lattice term, i.e., Eq. (\ref{eq:Saturation}) +
Eq. (\ref{eq:Bessel}), for three different \textit{ans\"{a}tze}: the vortex
beam with $m=1$, the azimuthon with $m=2$, and finally, an the elliptic
quadrupole. The two former \textit{ans\"{a}tze} carry the orbital angular
momentum due to their phase structure, which drives their clockwise rotation
in the course of the propagation, while preserving the overall intensity
shape, as shown in Fig. \ref{fig:Saturation_Bessel1}(a) and Fig. \ref{fig:Saturation_Bessel1}(b). It is
worthy to note that, in the pure saturable medium, an attempt to produce
asymmetric quadrupoles by means of the VA\ optimization does not produce any
robust mode, but, if the Bessel term (\ref{eq:Bessel}) is added, the
quadrupole develops into a noteworthy stable dynamical regime of periodic
transitions between vertical and horizontal dipole modes, as shown in Fig. %
\ref{fig:Saturation_Bessel1}(c).

\begin{figure}[th]
\centering
\includegraphics[trim={5.5cm 12.5cm 4.25cm
9cm},clip,width=0.8\textwidth]{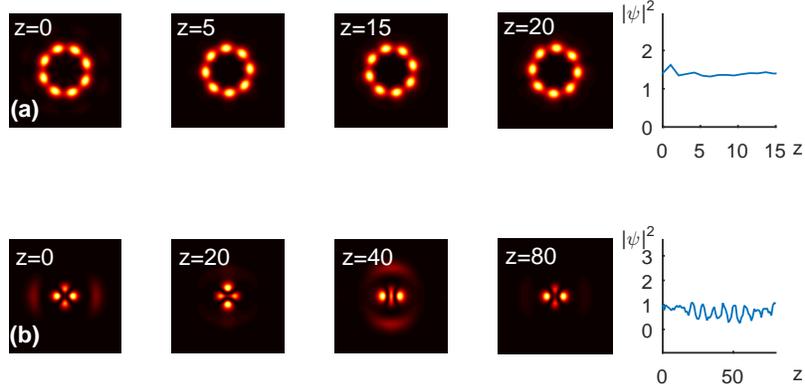}
\caption{The same as in Fig. \protect\ref{fig:Saturation_KB1}, but in
the medium with the combination of the nonlinear terms corresponding to the
combination of Eqs. (\protect\ref{eq:Saturation}) + (\protect\ref{eq:Bessel}%
), with $s=0.05$, $p=15$ and $k=1$. (a) The fourth-order azimuthon with $\protect%
\delta =\protect\pi /3$, $\mathbf{A_{1,2,3}}=(0.4216,2.8725,2.8725)$, and TDA $%
20\times 20$. (b) The asymmetric quadrupole with $\protect\epsilon =5\protect%
\pi /16$, $\mathbf{A_{1,2,3}}=(1.2722,2.2057,1.3310)$, and TDA $15\times 15$.}
\label{fig:Saturation_Bessel0}
\end{figure}

Finally, we address the propagation of an azimuthon with $m=4$ and
asymmetric quadrupole, under the combined action of the saturable
nonlinearity and zeroth-order Bessel lattice, as shown in Fig. \ref%
{fig:Saturation_Bessel0}(a) and Fig. \ref%
{fig:Saturation_Bessel0}(b), respectively. The orbital angular
momentum of the modes drives their rotation in the course of the
propagation.\ In particular, the former pattern is predicted in quite a
robust form, while the latter one is transformed into the above-mentioned
regime of periodic transitions between vertical and horizontal dipole modes.
Additional results produced by the general ansatz given by Eq. (\ref%
{Gen_Trial_function}) and the VA outlined above, in the combination with
direct simulations, will be reported elsewhere.
\begin{table}[th]
\caption{Anharmonic terms in the Lagrangian}
\label{tab:Nonlinear_potentials}\centering
%\vspace{-0.4cm}
\begin{tabular}{p{0.20\textwidth}p{0.75\textwidth}}
\vspace{-0.5cm} Kerr &
\vbox{\begin{equation}
\mathcal{NL}(|\psi|^2,\textbf{r})= -1/2 |\psi|^4,
\label{eq:Kerr}
\end{equation}} \\
\vspace{-0.5cm} Saturation & \vspace{-0.7cm}
\vbox{\begin{equation}
\mathcal{NL}(|\psi|^2,\textbf{r})=(\ln(s|\psi|^2+1)-s|\psi|^2)/s^2,
\label{eq:Saturation}
\end{equation}} \\
\vspace{-0.45cm} $n$-th order Bessel lattice & \vspace{-0.7cm}
\vbox{\begin{equation}
\mathcal{NL}(|\psi|^2,\textbf{r}) = - p \left[ J_n{\ (k\sqrt{x^2+y^2} )}\right]^2|\psi|^2.
\label{eq:Bessel}\vspace{-2cm}
\end{equation}}%
\end{tabular}%
\end{table}

\section{Conclusion}

In this work we report an efficient algorithm for full numerical treatment
of the variational method predicting two-dimensional solitons of GNLSE
(generalized nonlinear Schr\"{o}dinger equation) with various nonlinear
terms, which arise in nonlinear optics. A general class of the solitons is
considered, including vortices, multipoles, and azimuthons. The method
predicts solutions for the self-trapped beams which demonstrate robust
propagation in direct simulations. Further work with the use of the
algorithm is expected, making use of more complex flexible \textit{ans\"{a}%
tze}, which should improve the accuracy (making the calculations more
complex, which is not a critical problem for the numerically implemented
algorithm). In particular, it is planned to apply the VA to models of
nonlocal nonlinear media, where the Lagrangian integrals become quite
difficult for analytical treatment. Another promising direction is
application of the algorithm to three-dimensional settings, where the
analytical work is hard too.

\section*{Funding}
Consejo Nacional de Ciencia y Tecnolog\'{i}a (CONACYT) (243284); National Science Foundation (NSF) and Binational (US-Israel) Science Foundation (2015616); Israel Science Foundation (1286/17).

% We acknowledge financial support from CONACYT through grant \#243284. The
% work of B.A.M. was partly supported by the joint program in physics between
% NSF and Binational (US-Israel) Science Foundation through project No.
% 2015616, and by the Israel Science Foundation through Grant No. 1286/17

\end{document}